\documentclass[amsmath,amssymb,
 aps, prd, showkeys,
	nofootinbib,
	floatfix,
        superscriptaddress,]{revtex4-2}

\usepackage{placeins}
\usepackage{lineno}
\usepackage{graphicx}
\usepackage{dcolumn}
\usepackage{bm}

\usepackage[utf8]{inputenc}
\usepackage{aas_macros}
\pdfoutput=1
\usepackage{hyperref}

\usepackage[caption=false]{subfig}
\hypersetup{ 
			colorlinks=true,
			pdfauthor={Kyle Boone},
			pdftitle={WST TBD},
			citecolor=blue
			}

\usepackage{xcolor}
\usepackage{multirow}
\usepackage{array} 
\usepackage{ctable}

\begin{document}

\title{Beyond Linear Bias Expansions for AbacusSummit Halos at $z=8$}

\author{Kyle K.\ Boone}
\email{kboone@g.harvard.edu}
\affiliation{Department of Physics, Harvard University, Cambridge, MA 02138, USA}

\author{Daniel J.\ Eisenstein}
\email{deisenstein@cfa.harvard.edu}
\affiliation{Harvard-Smithsonian Center for Astrophysics, 60 Garden St., Cambridge, MA 02138, USA}

\date{\today}


\begin{abstract}
We study the non-Gaussianity of the large-scale clustering of high-redshift halos, seeking to assess which terms of standard bias expansions are needed to understand these highly biased populations.
We find that the clustering can be well modeled with only linear and quadratic bias parameters while assuming a Gaussian underlying matter field.
Our analysis focuses on AbacusSummit halos at redshift $z=8$.
We work with halos of mass at least $1\times10^{11}h^{-1}M_\odot$ in boxes of side length $2h^{-1}$Gpc.
Measurements of bias coefficients are made by fitting bias expansions to the halo power spectrum and bispectrum.
Tidal bias is not detected with only a $\sim0.1\sigma$ deviation from $0$, but we see a $17\sigma$ level detection for a bias term of the form $\delta^2$.
A bias term of the form $\delta^3$ is weakly detected at the $1.3\sigma$ level.
Nonlinear matter is also detected at a $1.3\sigma$ level.
To test how bias evolves, we run one test at $z=5$.
We use a mass threshold for halos that gives the same variance in the halo field as our $z=8$ sample.
Bias is smaller at $z=5$ and a tidal bias is detected at the $1\sigma$ level.
Bias coefficients at $z=5$ match a linear evolution of the $z=8$ bias coefficients to within $10\%$.
\newline
\end{abstract}

\maketitle
\twocolumngrid

\section{Introduction}
\label{sec:introduction}
Massive halos that can host luminous galaxies are rare at high redshift and only form in the most overdense regions of space.
As a result, the clustering of these objects is highly biased with respect to matter \citep{Kauffmann_1999, Jose_2017}.
While the clustering of halos is not directly observable, high-redshift galaxy clustering is able to confirm a large linear bias $b$ in the galaxy density field.
As examples, Lyman $\alpha$ emitters \citep{Ouchi_2017,Herrera_2025} show $b\sim4.5$ at $z=6.6$, Lyman $\alpha$ blobs \citep{Yang_2009,Yang_2010,Ramakrishnan_2023} show $b\sim7$ at $z=2.3$, Lyman break galaxies \citep{Barone_Nugent_2014, Harikane_2017, Hatfield_2018} show $b\sim8.6$ at $z=7.2$, and JWST observations \citep{Arita_2024, Dalmasso_2024, Paquereau_2025} show $b\sim9.6$ at $z=10.6$.

These bias values correspond to an amplitude of tracer density fluctuations similar to today's galaxies, with a standard deviation of the fractional mass fluctuations in $8h^{-1}$Mpc spheres $(\sigma_8)$ with values between $\sim$$0.5$ and $\sim$$2$.
Since any fractional overdensity field is mean $0$ with a minimum value of $-1$, any tracer with order unity fluctuations on a given filtering scale must have a probability distribution of fractional overdensity that is substantially skewed and hence must display non-Gaussian clustering properties on this scale.
For these high-redshift tracers, this implies non-Gaussianity for wavenumbers around $0.2h$ Mpc$^{-1}$.
This non-Gaussianity can enter the tracer field in two ways: the underlying matter field could have undergone a significant amount of nonlinear gravitational evolution, and the tracer field could have nonlinear dependencies on the matter field, which is known as a bias expansion.

At large scales, bias expansions are typically viewed as perturbative expansions in the matter overdensity \citep{Bias}, which is assumed to be $\ll 1$.
However, if the ratios of different bias coefficients is of the same order as the matter overdensity, the ordering of terms in this perturbative expansion can change.
In particular, a term in the bias expansion at a higher order in the matter overdensity than another term could be more significant in clustering if the ratio of their bias coefficients is large enough.
The large bias values in high-redshift tracers suggest that this could be the case, and if so an alternative expansion should be used.
In this work, we study the bias of halos produced in the AbacusSummit \citep{Abacus_Sim_Suite} suite of N-body simulations.
Our analysis focuses on halos at redshift $z=8$, although we perform tests at $z=5$ to see how the bias coefficients evolve.
We choose these redshift values to match the available Abacus halo catalogs.

To measure bias coefficients from halo catalogs, we use the power spectrum and the bispectrum of the halos.
We then construct all possible contributions to the power spectrum and bispectrum that arise from our bias expansion.
Each term is weighted by some combination of bias coefficients.
After producing models for all possible contributions, we are able to fit bias coefficients to the AbacusSummit data.
By performing these measurements on $25$ separate boxes, we are able to generate constraints for bias coefficients.

The rest of the paper is organized as follows: in Section \ref{sec:bias} we review the Eulerian bias expansion and show how to properly renormalize our bias operators.
We start Section \ref{sec:sims} by discussing the measurement of the power spectrum and bispectrum from the AbacusSummit halo catalogs.
We then cover the generation of the different bias operators and how these are used to generate our functionals for the power spectrum and bispectrum.
In Section \ref{sec:analysis} we discuss the fitting of the bias coefficients, the results of which are shown in Section \ref{sec:results}.
Finally, we conclude in Section \ref{sec:conclusion}.

\section{Bias Expansion}
\label{sec:bias}
This section will focus on developing a bias expansion for a general biased tracer.
In order to have robust clustering predictions that are insensitive to the small scale cutoff, we also introduce renormalized bias operators.
Throughout this section we will be switching between real space and Fourier space.
We use the following Fourier convention:

\begin{gather}
    f(\vec{k}) = \int d^3x \,\, f(\vec{x})\exp[i\vec{k}\cdot\vec{x}],\\
    f(\vec{x}) = \int \frac{d^3k}{(2\pi)^3} \,\, f(\vec{k})\exp[-i\vec{k}\cdot\vec{x}].
\end{gather}

For any density field $\rho$, we define the overdensity field as:

\begin{equation}
    \delta(\vec{x}) \equiv \frac{\rho(\vec{x})}{\bar{\rho}}-1,
\end{equation}
with $\bar{\rho}$ denoting the average density.

In this work, if no subscript is attached to $\delta$ it will denote the overdensity of all matter.
Any biased tracer of matter will also have an overdensity field.
Since this work focuses on halos we will denote this overdensity field as $\delta_h$.
The simplest ansatz to relate the two overdensities is with a linear bias:

\begin{equation}
    \delta_h(\vec{x})=b_1\delta(\vec{x}).
\end{equation}

This can be generalized to include other operators which we will define later:

\begin{equation}
    \delta_h(\vec{x}) = \sum b_O O(\vec{x}).
\end{equation}

For a comprehensive review of bias expansions, we refer the reader to \citet{Bias}.
Following their notation, we will refer to terms which are powers of the matter density ($\delta^n$) as local-in-matter-density (LIMD) terms.
In the following section we will briefly summarize the results necessary for this work.

\subsection{Eulerian Bias Expansion}

In this work we only consider what \citet{Bias} refers to as local bias operators, which are more general than just LIMD operators.
Local bias operators are constructed by taking exactly two derivatives of the gravitational potential $\Phi$ whenever it appears.
These operators are built from the matter density $\delta(\vec{x})\propto\nabla^2\Phi(\vec{x})$ and the tidal tensor, which is most easily defined in Fourier space:

\begin{equation}
    K_{ij}(\vec{k})\equiv\left[ \frac{k_ik_j}{k^2}-\frac{1}{3}\delta^K_{ij}\right]\delta(\vec{k}),\label{eq:tidal}
\end{equation}
where $\delta^K$ is the Kronecker delta.
To enter the bias expansion, the indices of $K_{ij}$ must be contracted.
By definition it is a traceless tensor, so the lowest order operator that includes the tidal tensor is $K^2\equiv(K_{ij})^2$, with the contraction occurring in real space.
Higher derivative terms are in general allowed in bias expansions, but we find no noticeable systematic errors in our clustering predictions when we do not include them.
To simplify our bias model in this work we do not include higher derivative terms.

The matter field itself ($\delta$) is an expansion in terms of the linear matter density $\delta^{(1)}$.
For an in-depth review, we refer the reader to \citet{PT_Review}.
For the purposes of this work, we will only need the leading order nonlinearity:

\begin{gather}
    \delta^{(2)}(\vec{k}) = \int \frac{d^3p}{(2\pi)^3}F_2(\vec{p},\vec{k}-\vec{p})\delta^{(1)}(\vec{p})\delta^{(1)}(\vec{k}-\vec{p}),\\
    F_2(\vec{p},\vec{q})\equiv\frac{5}{7}+\frac{1}{2}\frac{\vec{p}\cdot\vec{q}}{pq}\left(\frac{p}{q}+\frac{q}{p}\right)+\frac{4}{7}\frac{(\vec{p}\cdot\vec{q})^2}{p^2q^2}.
\end{gather}
From this point, we will drop the superscript $(1)$ from the linear matter density and write it as $\delta$.
All terms that appear in our final bias expansion besides $\delta^{(2)}$ use only the linear matter density.
For example, Eq. \ref{eq:tidal} uses only the linear matter density to construct the tidal field.
This will be fully justified in the next subsection.

We can now write out the density of any biased tracer as an expansion in the perturbative parameter $\delta$.
When working with high-redshift halos, there are other perturbative parameters that can be leveraged: ratios of the bias coefficients.
From a heuristic point of view, at high redshifts we expect halos to form only in the most overdense regions.
Higher powers of $\delta$ (such as $\delta^2$ and $\delta^3$) pick out overdense regions, and thus we expect large bias parameters for these terms.
Specifically, we would expect $b_3>b_2>b_1$, which is what we find in Section \ref{sec:results}.
If halos just form in the most overdense regions, then the local shape of the matter density does not matter, so tidal bias should be suppressed relative to other terms.
This assumption agrees with the work of \citet{Jeong_2009} in which high-redshift galaxy clustering was accurately modeled with no tidal bias.
When fitting for our bias coefficients, we find all of these heuristic assumptions to be accurate.
Therefore, although there are many terms which can enter the bias expansion at third order, in this work we only include $\delta^3$:

\begin{align}
    \delta_h(\vec{x})=\,&b_1\left([\delta(\vec{x})]+[\delta^{(2)}(\vec{x})]\right) + b_2 [\delta^2(\vec{x})] \nonumber\\+&b_{K^2}[K^2(\vec{x})] + b_3 [\delta^3(\vec{x})],
    \label{eq:bias_expansion}
\end{align}
with the brackets denoting renormalized quantities, which we turn to in the next subsection.
When calculating the power spectrum and bispectrum to fit our bias parameters, we include only leading order contributions from tidal terms and nonlinear matter.
We go to leading order both in $\delta$ and the bias coefficients.
For instance, when calculating the halo power spectrum we would include $b_1b_2\langle \delta^2\delta^{(2)}\rangle$ but not $b_1^2\langle\delta\delta^{(3)}\rangle$, which is why we only expand the matter density to second order.
We go one order beyond this when using only LIMD terms as their correlators will include the larger bias coefficients.

\subsection{Renormalized Bias Operators}
\label{sec:renorm}

Without applying any renormalization, Eq. \ref{eq:bias_expansion} has numerous problems.
We could have $\langle\delta_h\rangle\neq0$ since $\langle\delta^2\rangle\neq0$, which is not allowed by the definition of $\delta_h$.
A more subtle issue occurs when calculating the power spectrum of $\delta_h$.
Even if we redefine $\delta^2$ to be mean zero, we get the following contribution to the power spectrum:

\begin{equation}
    P_h(k)\supset 2\int\frac{d^3q}{(2\pi)^3}P(q)P(|\vec{k}-\vec{q}|).
\end{equation}
This indefinite integral extends to infinitely small scales where any cosmological theory cannot be expected to hold, so we necessarily must truncate it at some minimum scale $\Lambda$.
In practice, we achieve this by zeroing out our fields at scales smaller than $\Lambda$.

This scale $\Lambda$ can induce cutoff dependence even in the matter power spectrum.
This is addressed in the EFT of Large-Scale Structure \citep{Baumann_2012, Carrasco_2012, Pajer_2013, Carrasco_2014, philcox2025eft} by introducing counterterms.
In practice, the counterterm for $\delta^{(2)}$ is small \citep{philcox2025eft}, so we ignore it in this work.

Next we must renormalize all of our other bias operators \citep{McDonald_2006, McDonald_2009, Schmidt_2013}.
We follow the formalism from \citet{Renormalize} for this.
To renormalize an operator $O$, we add counterterms so that the following condition is met:

\begin{gather}
    \langle [O(\vec{k})]\delta(\vec{k}_1)\cdot\cdot\cdot \delta(\vec{k}_m)\rangle = \langle O(\vec{k})\delta(\vec{k}_1)\cdot\cdot\cdot \delta(\vec{k}_m)\rangle_\text{tree},\nonumber\\ \text{at }\vec{k}_i=0,\forall i,\label{eq:renorm_cond}
\end{gather}
where brackets denote the renormalized operator and ``tree" means the tree level result.
This becomes simple in our case where only leading order matter nonlinearities are used.
In the bispectrum, the only leading order contribution from matter nonlinearities comes from $b_1^3\langle\delta\delta\delta^{(2)}\rangle$.
In the power spectrum, the leading order contribution is $b_1b_2\langle \delta^2\delta^{(2)}\rangle$.
The only other contribution from a matter nonlinearity that would be at the same order both in bias coefficients and the linear matter density would be $b_1b_2\langle \delta^2\delta\rangle$ if the $\delta^2$ field included a nonlinearity.
However, our renormalization condition requires that this evaluates to $0$, which justifies constructing $\delta^2$ out of only the linear matter density.
Analogous justifications follow for the tidal field and $\delta^3$.
Finding counterterms to satisfy Eq. \ref{eq:renorm_cond} then becomes simple.
Our renormalized operators are as follows:

\begin{alignat}{2}
    &[\delta] &&= \delta,\\
    &[\delta^{(2)}] &&= \delta^{(2)},\\
    &[\delta^2] &&= \delta^2-\sigma(\Lambda)^2,\\
    &[K^2] &&= K^2 - \frac{2}{3}\sigma(\Lambda)^2,\\
    &[\delta^3] &&= \delta^3 - 3\delta\sigma(\Lambda)^2,
\end{alignat}
with $\sigma(\Lambda)^2$ denoting the variance of the field after smoothing at scale $\Lambda$. 
In following sections we will not include brackets as we will always be dealing with renormalized operators.

In practice the power spectrum and bispectrum can still have some dependence on $\Lambda$. 
The renormalization condition ensures that this dependence will vanish with the exception of a potential stochastic term \citep{Bias} when dealing with scales $k$ such that $k/\Lambda$ is small.
All of the scales used in this work satisfy this condition.
We add general noise terms as nuisance parameters as described in Section \ref{sec:sims} which will absorb any stochastic dependence.

\section{Simulated Observables}
\label{sec:sims}
We use the halo power spectrum and bispectrum as our simulated observables.
While halos are not directly observable in surveys, they are in the N-body simulations used in this work.
From an overdensity field $\delta(\vec{x})$, the power spectrum and the bispectrum respectively are defined as:

\begin{gather}
    (2\pi)^3 \delta^D(\vec{k}+\vec{k}') P(\vec{k}) = \langle\delta(\vec{k})\delta(\vec{k}')\rangle,\\
    (2\pi)^3 \delta^D(\Sigma_i\vec{k}_i) B(\vec{k}_1,\vec{k}_2,\vec{k}_3) = \langle\delta(\vec{k}_1)\delta(\vec{k}_2)\delta(\vec{k}_3)\rangle,
\end{gather}
where the Dirac delta functions ensures statistical homogeneity and $i$ in the second delta function runs from $1$ to $3$. 
Statistical isotropy reduces the dimension of the arguments further by requiring that $P$ can only depend on the magnitude of $\vec{k}$ and $B$ can only depend on the triangle configuration.
This means we can fully characterize the power spectrum by $P(k)$ and the bispectrum by $B(k_1, k_2, k_3)$, where the loss of the vector symbol implies we are looking at the magnitude of the wavenumbers only.

The above equations technically apply for an infinite volume limit. 
In our case where all halos are constrained to a box of volume $V$, we can adjust the equations by substituting $V$ in for $(2\pi)^3$.
Additionally, for a finite volume we replace the Dirac delta functions with Kronicker delta functions $\delta^K(\vec{k}+\vec{k}',0)$ and $\delta^K(\vec{k}_1+\vec{k}_2+\vec{k}_3,0)$ for the power spectrum  and bispectrum respectively.
We use the BSKIT package \citep{BSKIT,NBODYKIT,Corr_func_estimators,Bispec_theory_obs} to calculate all power spectra and bispecta in this work.
Since the bispectrum is symmetric with permutations of its arguments, we only calculate values where $k_3\leq k_2\leq k_1$.
To avoid the squeezed limit issues shown in \citet{Squeezed_Limit_Problems} we require $k_1\leq 0.9(k_2+k_3)$.

\subsection{Abacus Halo Catalogs}

We use the AbacusSummit \citep{Abacus_Sim_Suite} suite of N-body simulations to analyze halo clustering.
These were generated with the Abacus code \citep{Abacus_Code} run on the Summit supercomputer.
For this work, we use the ``base" resolution simulations.
These consist of $6912^3$ particles of mass $2.11\times 10^9h^{-1}M_\odot$ in a box with side length $2h^{-1}$Gpc.
All of the simulations begin at $z=99$ with initial conditions given by second-order Lagrangian Perturbation Theory.
The simulations used in this work are then evolved using a cosmology corresponding to \textit{Planck} 2018 $\Lambda$CDM results \citep{Planck_Values}.
A total of $25$ boxes are produced with this cosmology, and these are the boxes from which we select halos.

Halo finding in AbacusSummit is performed with the CompaSO algorithm \citep{Compaso}.
The halos we use are found using a competitive spherical overdensity algorithm to find all particles inside a density threshold of $\Delta=200$.
These are referred to as ``L$1$" halos.
Our data consists of the positions of all halos above some desired mass threshold.
With our halos selected, the number density field of halos is:

\begin{equation}
    \rho_h(\vec{x}) = \sum_i \delta^D(\vec{x} - \vec{x}_i),
\end{equation}
where $\vec{x}_i$ is the position of the $i^{th}$ halo.
To obtain a more manageable density field, we use the Triangular Shaped Cloud (TSC) method to paint our halos onto a mesh.
To suppress the impacts of aliasing \citep{Alias}, we paint our halos onto a $2048^3$ mesh, leading to a Nyquist frequency of $k_{N}\approx3.2h$ Mpc$^{-1}$.
The highest frequency modes we use in the power spectrum or bispectrum are $k\approx0.4h$ Mpc$^{-1}$.
The resulting overdensity field will then be well approximated by $W(k)\delta_h(\vec{k})$, where $\delta_h(\vec{k})$ is the true overdensity field and $W(\vec{k})$ is given by \cite{Alias}:

\begin{equation}
    W(k) = 1-\sin^2{(\frac{\pi k}{2k_N})} + \frac{2}{15}\sin^4{(\frac{\pi k}{2k_N})}.
\end{equation}

While the impact of $W(k)$ is only $\sim4\%$ at the scales considered, we still take the effect into account when fitting for bias parameters.
After producing the overdensity field on the $2048^3$ grid, we smooth it to a $256^3$ grid to improve runtimes.
This smoothing is done by removing all Fourier modes with a higher frequency than the Nyquist frequency for the $256$ grid before inverse Fourier transforming back.
Since we only ever calculate out Fourier space correlators, this does not impact our observables, it just makes them quicker to calculate by reducing the memory load of the overdensity grids.

With these overdensity grids, we calculate both the power spectrum and bispectrum using BSKIT \citep{BSKIT}.
The power spectrum is calculated in bins of size $\Delta k = k_F \approx 0.003 h$ Mpc$^{-1}$.
The bins are centered on integer multiples of $k_F$ and range from $k_F$ to the Nyquist frequency of the $256$ grid.
The bispectrum is calculated by first finding all valid triangle configurations.
In this work, we select $\Delta k = 2k_F$ and center our bins on values between $1.5k_F$ and $63.5k_F\approx k_N/2$ for the $256$ grid Nyquist frequency.
These bins are used for $k_1$, $k_2$, and $k_3$.
Once all of the valid triangle configurations are sorted into these bins, the average of $\delta_h(\vec{k}_1) \delta_h(\vec{k}_2) \delta_h(\vec{k}_3)$ within the bin is used to measure the bispectrum.

\subsection{Simulated Bias Functionals}

To measure bias parameters, we first need to generate the different fields used in the bias expansion.
First, we construct a realization of $\delta$ using a linear power spectrum produced by CLASS \citep{CLASS} evolved to our redshift of interest.
When constructing $\delta(\vec{k})$, we fix the magnitude to match the linear power spectrum exactly, only allowing the phase to vary.
This is done to reduce noise on future power spectrum and bispectrum measurements.
We do a separate test allowing the magnitude of $\delta(\vec{k})$ to be distributed according to a Gaussian.
We find that fixing the magnitude produces no noticeable bias in the power spectrum or bispectrum and significantly reduces shot noise.

As required by EFT, before constructing other fields we first smooth the linear density field by applying a sharp $k$ cutoff

\begin{equation}
    \delta(\vec{k})\rightarrow \delta_\Lambda(\vec{k}) = \delta(\vec{k}) \Theta(\Lambda-k),
\end{equation}
where $\Theta$ is the Heaviside step function.
We let $\Lambda$ be $\sim1.6h$ Mpc$^{-1}$.
With this choice we are able to construct our fields without worrying about aliasing effects and $k/\Lambda$ is small for all scales probed with our observables.

As described in Section \ref{sec:bias}, we use $\delta_\Lambda$ to construct the following fields which we use in our bias expansion: $\delta^2_\Lambda$, $\delta^3_\Lambda$, $K^2_\Lambda$, $\delta^{(2)}_\Lambda$.
These fields are constructed out of the smoothed linear matter density before being smoothed themselves at the same scale $\Lambda$.
To avoid aliasing effects, we calculate $\delta^3$ (before smoothing it) as:

\begin{equation}
    \delta^3(\vec{x}) = \delta_\Lambda(\vec{x}) \delta_\Lambda^2(\vec{x}).
\end{equation}

All fields are renormalized as described in Section \ref{sec:renorm}.
We compress the fields down to grids of $256^3$ cells to improve efficiency using the same process as with the halo fields.
We use these fields as inputs to BSKIT to calculate the power spectra and bispectra contributions from different bias expansion fields.

\begin{figure*}
    \centering
     \includegraphics[width=0.90\textwidth]{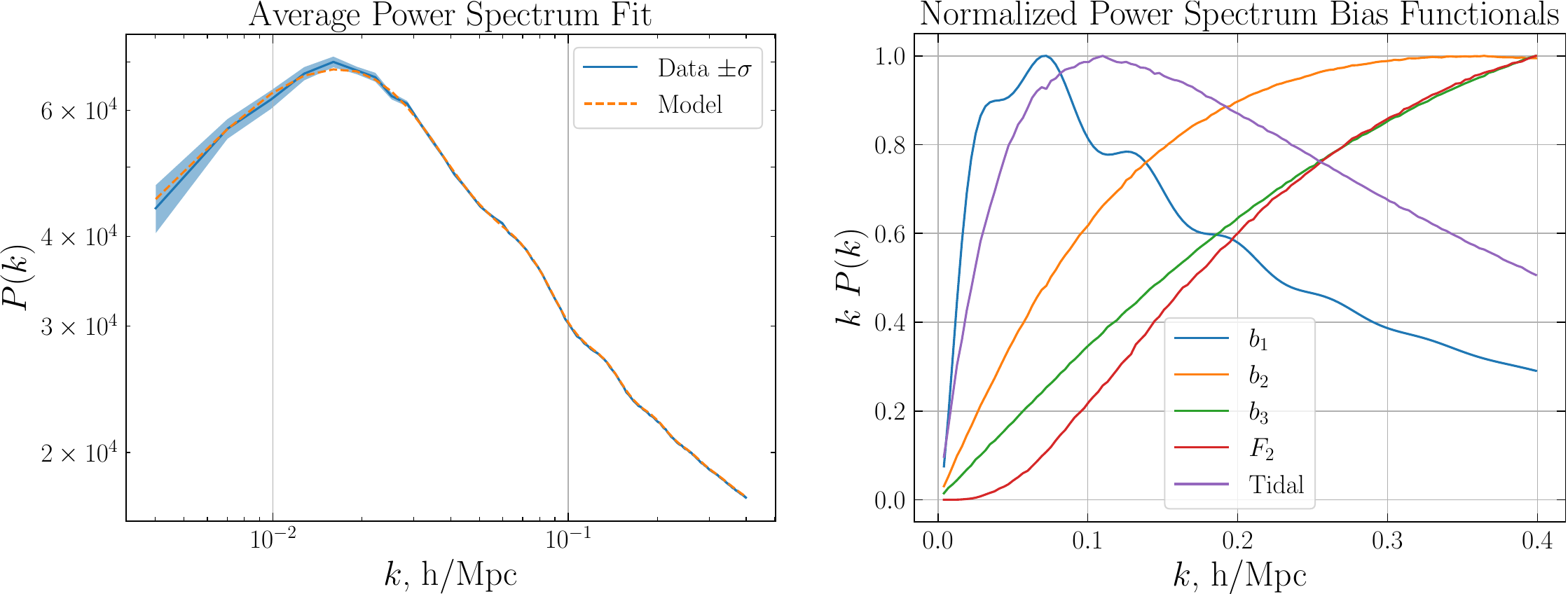}
      \caption{Left: Power spectrum data with standard deviation compared to the best fit. Our data is the average of the $25$ power spectrum realizations. The model is the average of our $25$ fits, and it shows no significant deviations from the data. Right: The bias functionals that appear in Eq. \ref{eq:power_functionals} in the order listed there. For the LIMD terms, higher-order operators peak at smaller scales and look more similar to a constant shot noise term.} 
      \label{fig:Power_Plot}
\end{figure*}

As mentioned in Section \ref{sec:bias}, we will work to leading order both in $\delta$ and bias coefficient ratios in matter nonlinearities and tidal terms.
When using only LIMD terms, we will go one order beyond this.
With that in mind, the power spectrum will have contributions from the following correlations, which are all shown on the right side of Figure \ref{fig:Power_Plot}:

\begin{equation}\label{eq:power_functionals}
    \langle\delta_\Lambda\delta_\Lambda\rangle,\,\,\,\,
    \langle\delta_\Lambda^2\delta_\Lambda^2\rangle,\,\,\,\,
    \langle\delta_\Lambda^3\delta_\Lambda^3\rangle,\,\,\,\,
    \langle\delta_\Lambda^2\delta_\Lambda^{(2)}\rangle,\,\,\,\,
    \langle\delta_\Lambda^2K_\Lambda^2\rangle.
\end{equation}

We follow the same procedure when determining which terms to use in the bispectrum where matter nonlinearities and tidal terms are only included at leading order.
The bispectrum will have contributions from the following correlations:

\begin{gather}
    \langle\delta_\Lambda\delta_\Lambda\delta_\Lambda^2\rangle,\,\,\,\,
    \langle\delta_\Lambda\delta_\Lambda\delta_\Lambda^{(2)}\rangle,\,\,\,\,
    \langle\delta_\Lambda\delta_\Lambda K_\Lambda^2\rangle,\nonumber\\
    \langle\delta_\Lambda\delta_\Lambda^2\delta_\Lambda^3\rangle,\,\,\,\,
    \langle\delta_\Lambda^2\delta_\Lambda^2\delta_\Lambda^2\rangle,
\end{gather}

\begin{figure*}
    \centering
     \includegraphics[width=0.90\textwidth]{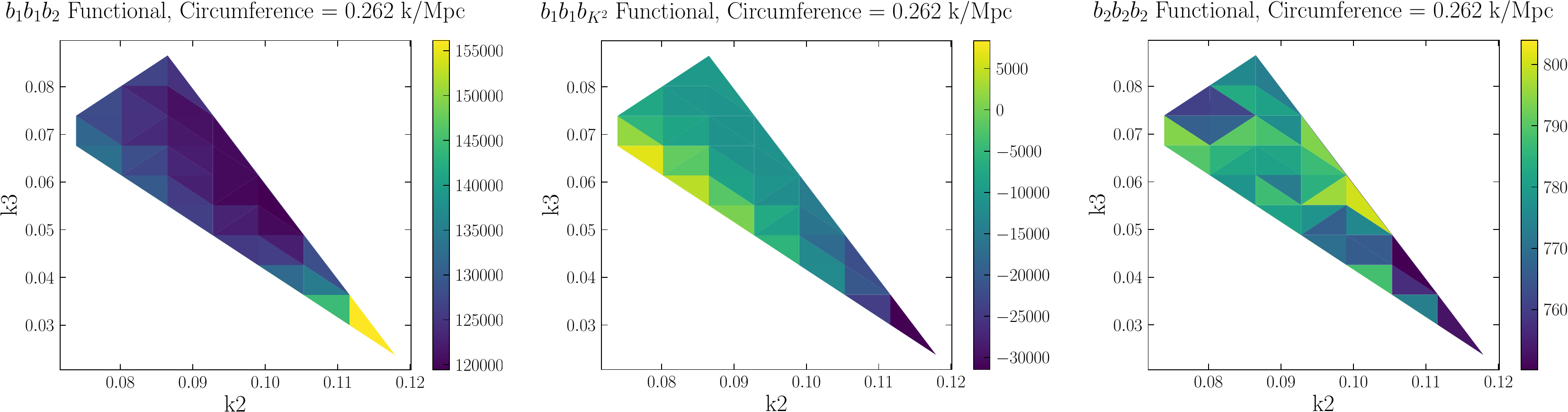}
      \caption{Left: The bispectrum functional corresponding to $b_1b_1b_2$. The bispectrum is plotted as a function of $k_2$ and $k_3$ with the sum $k_1+k_2+k_3$ being fixed. The squeezed limit is on the bottom right, while the equilateral limit is at the top. Middle: The bispectrum functional corresponding to $b_1b_1b_{K^2}$. Right: The bispectrum functional corresponding to $b_2b_2b_2$. All three functionals have visually distinct patterns, which suggests that we will be able to use them to fit for bias coefficients.} 
      \label{fig:Bispec_Functionals}
\end{figure*}

Figure \ref{fig:Bispec_Functionals} shows the functionals for the $b_1b_1b_2$, $b_1b_1b_{K^2}$, and $b_2b_2b_2$ terms.
Our bispectrum plots are shown as functions of $k_2$ and $k_3$ for a fixed ``circumference", $k_1+k_2+k_3$.
Distinct patterns can be seen for each of the functionals.
The functional for $b_1b_1b_2$ peaks in the limit where $k_3\rightarrow 0$ and decays away from this limit.
Alternatively, the tidal functional shows a peak when $k_2\approx k_3$ which decays away from this point and actually becomes strongly negative in the limit where $k_3\rightarrow 0$.
The functional for $b_2b_2b_2$ shows the least clear pattern, but it is immediately clear from the colorbar that this functional appears at a higher order in classic perturbation theory since it is suppressed by over an order of magnitude compared to the other two functionals plotted.
Compared to the other functionals it is also the most uniform.

We calculate all power spectrum and bispectrum terms listed above from $25$ separate generated sets of fields.
The average across these realizations is used when fitting to halo data.
All of these bispectrum contributions are calculated using a version of BSKIT which was modified to be able to handle cross correlations.
When comparing these simulated power spectra and bispectra to the halo data from AbacusSummit, we need to account for the fact that the halos are discrete.
The finite number of halos will have an impact on the power spectrum and bispectrum given by \citep{Abacus_Code}:

\begin{align}
    P_d(k) = &P(k)+\frac{1}{\bar{n}},\\
    B_d(k_1,k_2,k_3) = &B(k_1,k_2,k_3)\nonumber\\ + &\frac{1}{\bar{n}}(P(k_1)+P(k_2)+P(k_3)) + \frac{1}{\bar{n}^2},\label{eq:bi_noise}
\end{align}
where a subscript $d$ denotes a discrete correlator.
This impact is taken into account when fitting for our bias parameters, which we turn to next.

\section{Analysis}
\label{sec:analysis}
To fit for bias parameters, we perform a $\chi^2$ fitting. 
When performing our fits, we allow for some variation in the noise contributions. 
This can be understood as the impact of higher order bias expansion parameters or as the stochastic dependence on our smoothing scale.
We also account for the symmetry factors in all cross correlations.
For notational purposes, we will assume that all such symmetry factors are absorbed into correlation functions, which leads to us writing the following power spectrum expansion:

\begin{align}
    \tilde{P}_h(k,\vec{b}) = &b_1^2 \langle\delta_\Lambda\delta_\Lambda\rangle + b_2^2 \langle\delta^2_\Lambda\delta^2_\Lambda\rangle + b_3^2 \langle\delta^3_\Lambda\delta^3_\Lambda\rangle\nonumber\\ + &b_1b_2 \langle\delta^2_\Lambda\delta^{(2)}_\Lambda\rangle + b_2b_{K^2} \langle\delta^2_\Lambda K^2_\Lambda\rangle + b_0\frac{1}{\bar{n}}.\label{eq:power}
\end{align}

In the above we are dropping the $(2\pi)^3\delta^D(\vec{k}+\vec{k}')$ and the arguments of the bias expansion fields for the sake of brevity.
The parameter $b_0$ allows for the noise levels to change, although we would expect it to be $\sim$$1$.
With only $25$ halo catalog realizations, the covariance matrix of the power spectrum would be singular.
In order to produce the invertible covariance matrix necessary for a $\chi^2$ fit, we bin the power spectrum by $k$ into $10$ log spaced bins. 
With $P_{h,i}(k)$ being the $i^{th}$ halo catalog's power spectrum, we have the following:

\begin{equation}
    \chi^2_{P,i} = (\tilde{P}_h(k,\vec{b})-P_{h,i}(k))^TC^{-1}_P (\tilde{P}_h(k,\vec{b})-P_{h,i}(k)),
\end{equation}
where $C^{-1}_P$ is the inverse covariance matrix of the power spectrum.
Minimizing this will give bias fit parameters for each of the halo catalog realizations.
These would be fits based only on the power spectrum though, so if we want to use information from the bispectrum we must calculate a corresponding $\chi^2$ value for it.

Motivated by observations in the power spectrum noise level being largely degenerate with higher order contributions, we also allow the bispectrum noise levels to be variable.
With that in mind, after absorbing symmetry factors into correlators and dropping $(2\pi)^3\delta^D(\vec{k}_1+\vec{k}_2+\vec{k}_3)$ as well as arguments in the bias expansion fields, we get the following expansion:

\begin{align}
    \tilde{B}_h(&k_1,k_2,k_3,\vec{b})=  b_1^2b_2\langle\delta_\Lambda\delta_\Lambda\delta_\Lambda^2\rangle + b_1b_2b_3\langle\delta_\Lambda\delta_\Lambda^2\delta_\Lambda^3\rangle \nonumber\\ + &b_2^3\langle\delta_\Lambda^2\delta_\Lambda^2\delta_\Lambda^2\rangle + b_1^3 \langle\delta_\Lambda\delta_\Lambda\delta_\Lambda^{(2)}\rangle + b_1^2b_{K^2} \langle\delta_\Lambda\delta_\Lambda K_\Lambda^2\rangle \nonumber \\
    +&b_0'\frac{1}{\bar{n}} (\tilde{P}_h(k_1,\vec{b})+\tilde{P}_h(k_2,\vec{b})+\tilde{P}_h(k_3,\vec{b})) + b_0''\frac{1}{\bar{n}^2}. \label{eq:bispectrum}
\end{align}
Two additional parameters are added to allow for flexibility in the noise: $b_0'$ and $b_0''$.
These parameters add as much flexibility into the scaling of Eq. \ref{eq:bi_noise} as is possible with the symmetries of the bispectrum.
As with the power spectrum, the full covariance matrix of the bispectrum is singular.
The bispectrum having three arguments makes defining a binning scheme more complicated however.
Instead, we approximate the covariance as its diagonal estimated from the $25$ simulations.
For the smaller scales used in this analysis, this approximation ignores significant correlations.
We do not attempt to correct for this effect since when we compare our bias parameters found from the power spectrum alone and the bispectrum alone, we find no significant shifts.
With this diagonal covariance, we have the following:

\begin{align}
    \chi^2_{B,i} =& (\tilde{B}_h(k_1,k_2,k_3,\vec{b})-B_{h,i}(k_1,k_2,k_3))^T\nonumber\\&C^{-1}_B (\tilde{B}_h(k_1,k_2,k_3,\vec{b})-B_{h,i}(k_1,k_2,k_3)),\\
    \chi^2_i =& \chi^2_{P,i} + \chi^2_{B,i} \label{eq:final_chi2}
\end{align}.

This final equation implies that the bispectrum and power spectrum do not have significant cross correlations, which we assume in this work.
This final $\chi^2_i$ is minimized as a function of the bias parameters using the Optax \citep{Optax} package.
This is done for each halo catalog realization to find our best fit bias parameters.
We turn to these fits now.

\section{Results}
\label{sec:results}
\begin{figure*}
    \centering
     \includegraphics[width=0.90\textwidth]{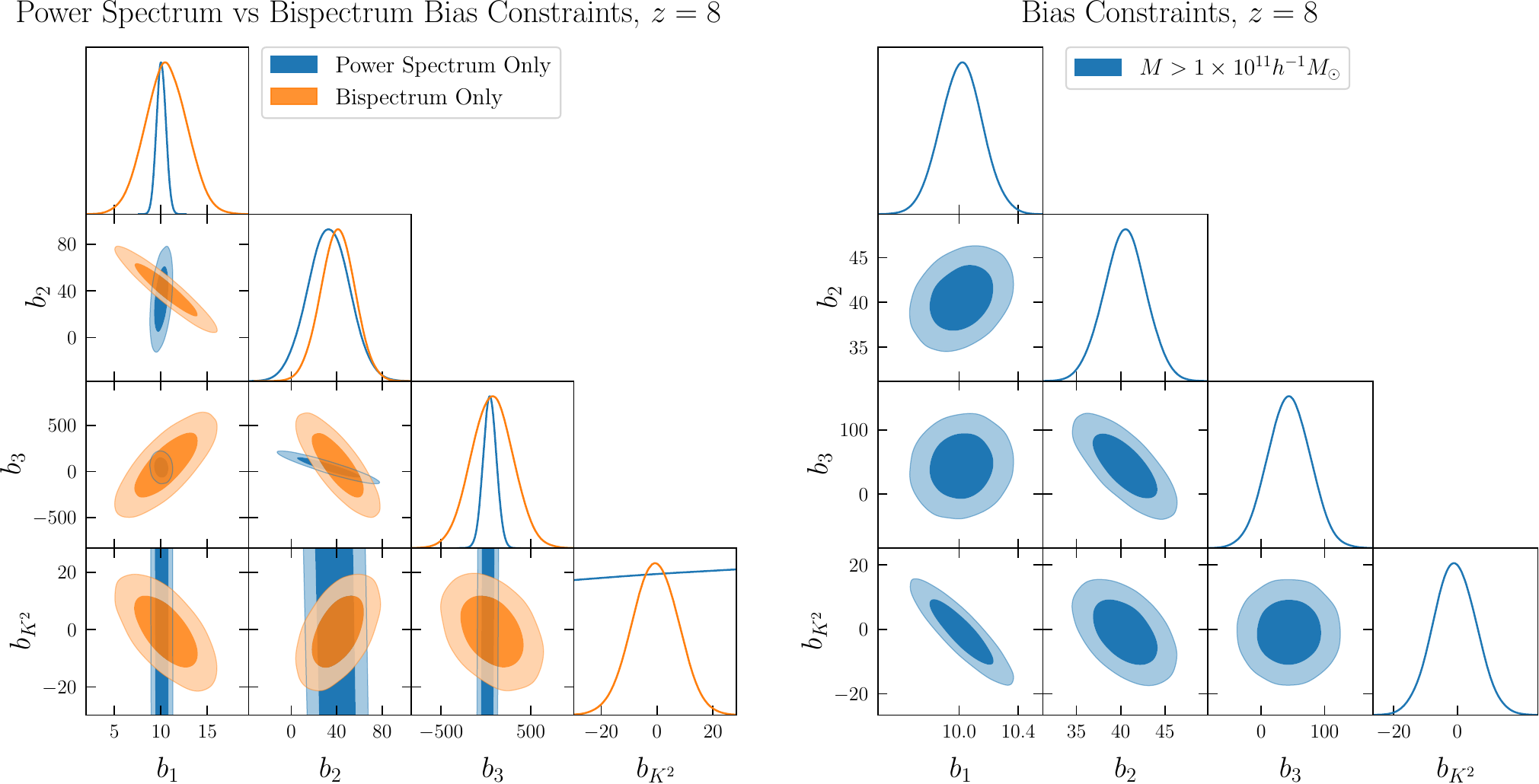}
      \caption{Left: constraints on the bias of halos with mass at least $10^{11}h^{-1}M_\odot$ at redshift $z=8$. The power spectrum and bispectrum are used to generate separate constraints. The constraints from both are consistent and show different degeneracies, which motivates combining them. Right: constraints on the same halos when the power spectrum is combined with the bispectrum. This is our baseline result, and as expected we see $b_1<b_2$ with a highly suppressed tidal bias. The uncertainty of $b_3$ makes it difficult to compare it directly with $b_2$.} 
      \label{fig:z8_Constraints}
\end{figure*}

\begin{table*}
\centering

\makebox[\textwidth]{
\begin{tabular}{|l|c|c|c|c|c|}
\hline
\multicolumn{6}{|c|}{\textbf{Bias Coefficient Constraints: Mean Value $\pm1\sigma$}} \\
\hline
Redshift, Mass Threshold & $b_1$ & $b_2$ & $b_3$ & $b_{K^2}$ & $b_{F2}$ \\
\hline\hline
$z=8$, $1\times10^{11}h^{-1}M_\odot$, Power                & $10.0\pm0.5$ & $30\pm20$ & $40\pm70$ & $100\pm300$ &  \\
\hline
$z=8$, $1\times10^{11}h^{-1}M_\odot$, Bispectrum           & $10\pm2$     & $40\pm10$ & $100\pm200$ & $-1\pm8$    &  \\
\hline
$z=8$, $1\times10^{11}h^{-1}M_\odot$                       & $10.0\pm0.1$ & $40\pm2$  & $40\pm30$ & $-1\pm7$    &  \\
\hline
$z=8$, $7\times10^{10}h^{-1}M_\odot$                       & $9.3\pm0.1$  & $33\pm1$  & $40\pm20$ & $-2\pm3$    &  \\
\hline
$z=8$, $1\times10^{11}h^{-1}M_\odot$, Variable $F_2$       & $10.0\pm0.3$ & $40\pm3$  & $50\pm40$ & $0\pm10$    & $1.1\pm0.9$ \\
\hline
$z=5$, $6.34\times10^{11}h^{-1}M_\odot$                    & $7.6\pm0.1$  & $20\pm1$  & $17\pm9$ & $-2\pm2$    &  \\
\hline
\end{tabular}
} 

\caption{$1\sigma$ constraints for all bias coefficients for all of our tests. Besides the first two rows, all constraints come from a combination of the power spectrum and the bispectrum. Since $b_{F2}$ is only a bias parameter if we allow for the $F_2$ kernel to be changed, it is not a bias coefficient for any of our tests besides the variable $F_2$ test.}
\label{tab:1_sig_constraints}
\end{table*}

We get $25$ bias data vectors, one from each of the halo catalog realizations.
We can use these to generate a covariance matrix of the bias coefficients to estimate their uncertainties from individual $2h^{-1}$Gpc boxes.
We use GetDist \citep{getdist} to plot these constraints.

For a first test to ensure that our bispectrum constraints are not biased with respect to the power spectrum, we look at constraints from each observable individually.
In this test, we look at halos with mass at least $10^{11}h^{-1}M_\odot$ at redshift $z=8$.
Our boxes have on average $6.6\times 10^5$ such objects, which gives a comoving number density of $8.2\times 10^{-5}h^3$Mpc$^{-3}$.
Our results are shown on the left side of Figure \ref{fig:z8_Constraints}.
Additionally, $1\sigma$ constraints for all bias coefficients for all of our tests are presented in Table \ref{tab:1_sig_constraints}.
The power spectrum and bispectrum constraints are consistent with each other, and they also have different degeneracies for some of the bias coefficients, making them promising as complementary observables.
The power spectrum gives tighter constraints on $b_1$ which is intuitive since at very large scales the halo power spectrum should look like the linear matter power spectrum multiplied by $b_1^2$.
The power spectrum also gives tighter constraints on $b_3$ while the bispectrum gives significantly tighter constraints on the tidal bias.

Since constraints from the power spectrum and bispectrum are consistent with each other, we combine the two observables (as in Eq. \ref{eq:final_chi2}) for all of our future fits.
For a general sense of the behavior of bias at high redshift, we first look at bias constraints for halos with mass at least $10^{11}h^{-1}M_\odot$ at redshift $z=8$ (the same criteria as in the previous paragraph).
Our results are shown on the right side of Figure \ref{fig:z8_Constraints}.
As expected, we note that $b_2>b_1$ while $b_{K^2}$ is consistent with $0$ within $1\sigma$.
We find that for all our tests $b_3\approx b_2$ within uncertainties.

\begin{figure*}
    \centering
     \includegraphics[width=0.90\textwidth]{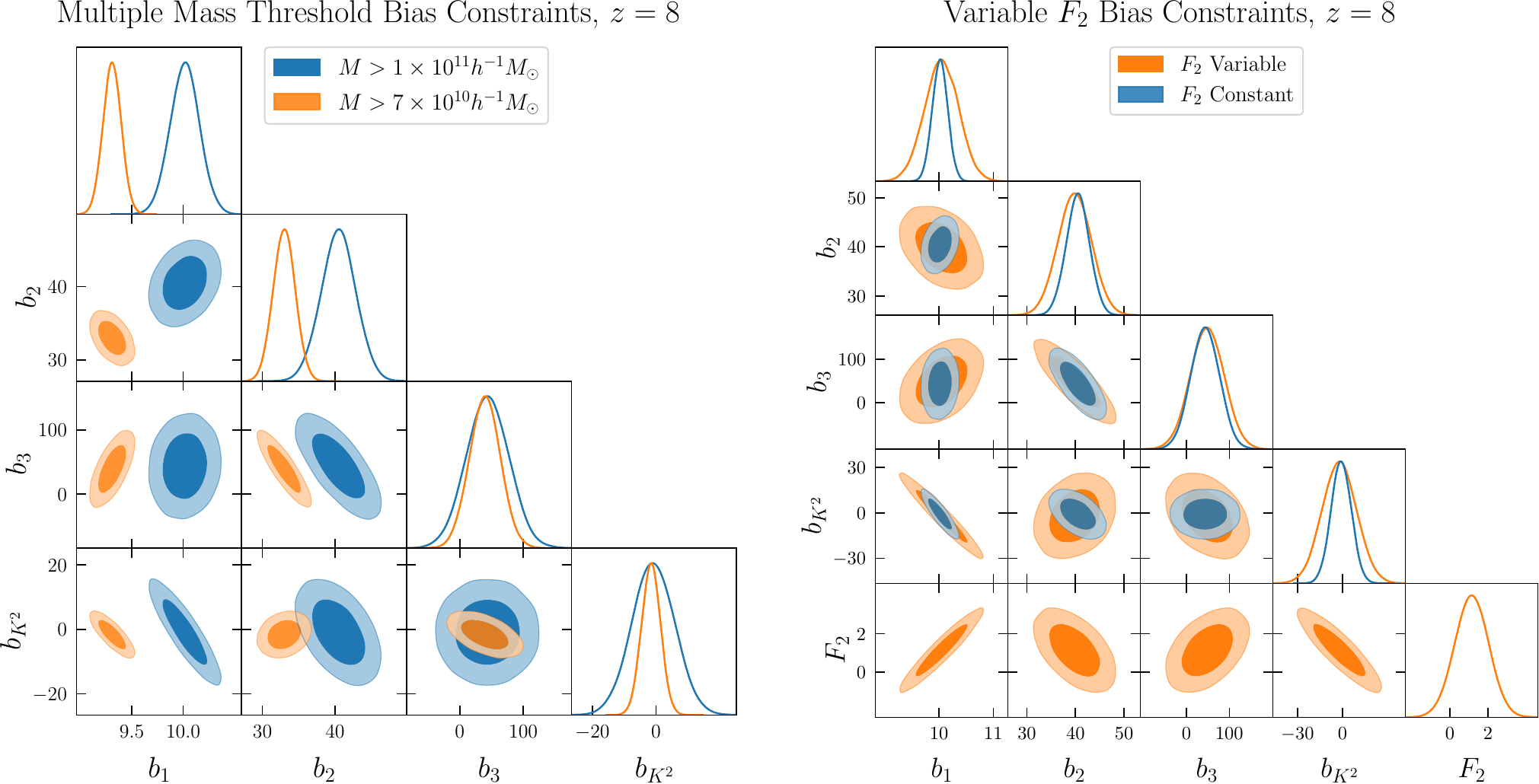}
      \caption{Left: bias constraints of halos with different mass thresholds at redshift $z=8$. When a lower mass is used there are more halos and they can form in less extreme environments. This leads to a reduction in the mean and variance of bias values. Right: bias constraints of halos with mass at least $10^{11}h^{-1}M_\odot$ at redshift $z=8$. An additional bias parameter $b_{F2}$ (denoted in the figure simply as $F_2$) is added to test the importance of matter nonlinearities. All bias values are consistent, and nonlinear matter is only weakly detected at a $\sim$$1\sigma$ level.} 
      \label{fig:Change_Constraints}
\end{figure*}

\begin{figure*}
    \centering
     \includegraphics[width=0.90\textwidth]{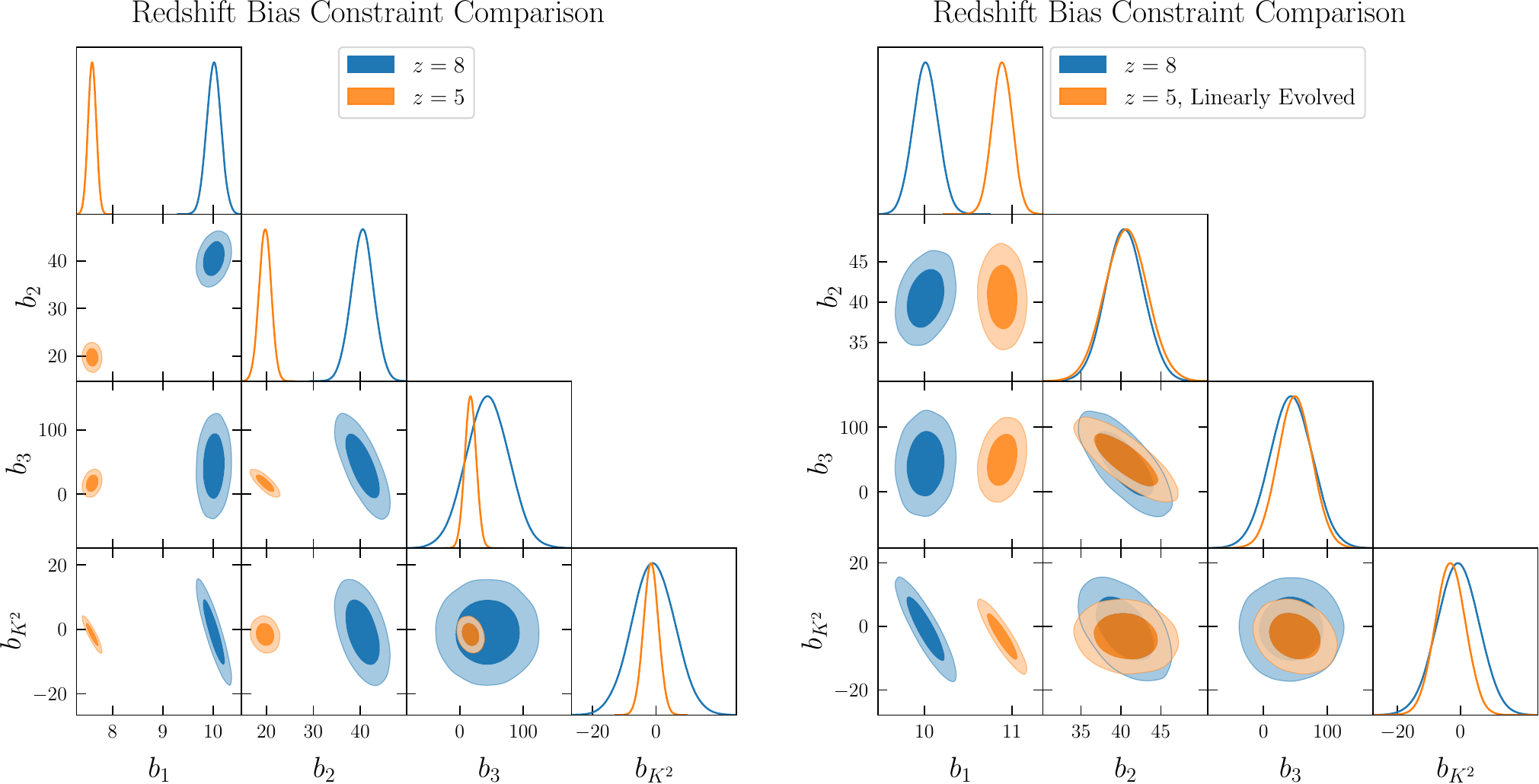}
      \caption{Left: bias constraints of halos at different redshifts. Mass thresholds are selected so that the halo fields have the same variance. Since the matter field overdensities have grown from $z=8$ to $z=5$, bias values are reduced. Right: the same as the left but with the $z=5$ bias constraints linearly evolved to $z=8$ assuming matter domination. The high consistency suggests that we are likely measuring the same objects at the two different redshifts.} 
      \label{fig:Redshift_Constraints}
\end{figure*}

Staying at redshift $8$, we additionally test how bias changes when the halo mass threshold is lowered. 
To test this, we look at bias constraints while using all halos in AbacusSummit, which corresponds to a mass threshold of roughly $7\times10^{10}h^{-1}M_\odot$.
Our boxes have on average $1.5\times 10^6$ such objects, which gives a comoving number density of $1.9\times 10^{-4}h^3$Mpc$^{-3}$.
Including more halos should have two effects.
First, the lower mass halos need less extreme conditions in order to form, so bias should tend to drop.
Second, more halos means more constraining power, so the bias coefficients across the $25$ simulations should become more consistent.
Both of these effects can be seen on the left side of Figure \ref{fig:Change_Constraints}.

One interesting question is whether the nonlinearity of matter is actually important.
So far we've seen that terms which are local in the matter density are detected with much higher significance than the tidal bias.
Similarly, we could wonder if the halo power spectrum and bispectrum actually require the matter to have nonlinearities.
To test for this, we add an additional bias parameter $b_{F2}$ anywhere $\delta_\Lambda^{(2)}$ appears.
This is not a true bias parameter since its value should be fixed to $1$, but it allows us to test how important the matter nonlinearities are.
This test is done with a mass cutoff of $10^{11}h^{-1}M_\odot$, and our results are shown on the right side of Figure \ref{fig:Change_Constraints}.
We find that $b_{F2}$ is consistent with a value of $1$ and is detected with approximately the same significance as $b_3$.
This result suggests that nonlinearities in matter are more important than tidal bias when it comes to describing biased tracers at high redshift.
This enforces the idea that the magnitude of local matter overdensities is much more important when forming early halos than the shape of the matter density.

We turn next to investigating how these behaviors shift at slightly lower redshifts.
To do this, we look at the the halo catalogs generated at $z=5$.
We compare bias coefficients to those found with a $10^{11}h^{-1}M_\odot$ mass cutoff at $z=8$.
At redshift $z=5$ we use a mass threshold of $6.34\times10^{11}h^{-1}M_\odot$.
This threshold is chosen so that the $z=5$ and $z=8$ halo overdensity fields have approximately the same variance.
This also corresponds to the two fields having approximately the same number of halos (to within $10\%$) as our boxes have on average $7.1\times 10^5$ such objects, which gives a comoving number density of $8.9\times 10^{-5}h^3$Mpc$^{-3}$.
For some intuition for what we expect to see here, we can assume that to first order, the halos in the two different selections are the same halos just evolved to different redshifts.
If the halos stay stationary over time, the two halo catalogs would have the same power spectrum and bispectrum while the underlying matter field continued to evolve.
Since the underlying matter field would have a higher variance at lower redshifts, we would expect to see this counteracted by lower bias coefficients at lower redshifts, which is exactly what we see on the left side of Figure \ref{fig:Redshift_Constraints}.
In particular, if we assume that the matter field is growing linearly in a period of matter domination and that we are tracking the same halos as they follow the gravitational potential, the linear bias at $z=8$ should be the linear bias at $z=5$ multiplied by a factor of $\sim$$1.43$ \citep{Fry_1996}.
The evolution of higher order bias coefficients due to gravitational evolution is more challenging to solve analytically, so we assume that the bias coefficients at $z=8$ are roughly the coefficients at $z=5$ multiplied by $1.43$ for every power of the matter field the bias coefficient multiplies.
We show the bias coefficients at $z=8$ compared to these rescaled $z=5$ coefficients on the right side of Figure \ref{fig:Redshift_Constraints}.
This shows that for $b_2$, $b_3$, and $b_{K^2}$ our simple rescaling gives the correct results.
For $b_1$ our results are similar with around a $10\%$ discrepancy, but since $b_1$ has by far the tightest error bars, the simple rescaling does not produce consistent results to within uncertainty.

Our power spectrum data and fit are shown on the left side of Figure \ref{fig:Power_Plot}.
This is for the $z=8$ halos with a mass cutoff of $10^{11}h^{-1}M_\odot$.
Our model fits the data well with no clear systematic errors.
The right of Figure \ref{fig:Power_Plot} shows the power spectrum functionals that appear in Eq. \ref{eq:power_functionals}.
Looking at the first three curves (these correspond to $\langle\delta^n\delta^n\rangle$ for $n$ between $1$ and $3$) we can see that larger powers of the matter field lead to power spectrum peaks at larger $k$. 
Larger powers of the matter field also lead to contributions that look increasingly like a pure noise contribution, which justifies our addition of extra bias parameters to rescale the noise.

\begin{figure*}
    \centering
     \includegraphics[width=0.90\textwidth]{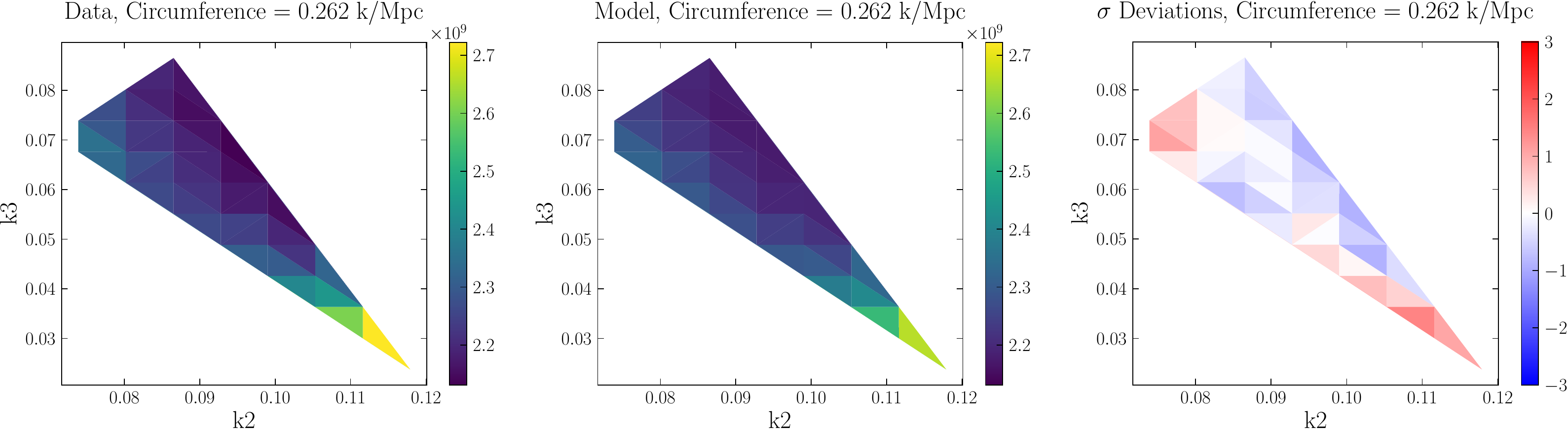}
      \caption{Left: Bispectrum data averaged over the $25$ realizations. Plotting in terms of $k$ values is done as in Figure \ref{fig:Bispec_Functionals}. Middle: The average model fit for the bispectrum. Right: The difference of the data and model bispectrum divided by the standard deviation at each point. All bispectra configurations shown are within $3\sigma$, showing that our model is able to accurately capture the behavior of the bispectrum.} 
      \label{fig:Bispec_Plot}
\end{figure*}

Turning to the bispectrum for the same set of objects, our fit and data are shown at fixed $k_1+k_2+k_3$ in Figure \ref{fig:Bispec_Plot}.
The rightmost plot shows that there are not clear systematic errors in our fit and that most residuals are within $2\sigma$ as we expect.
While we only show one circumference here for brevity, general patterns are consistent across circumferences.

\section{Conclusion}
\label{sec:conclusion}

Studies of high-redshift galaxies consistently find bias values large enough to require significantly non-Gaussian clustering for wavenumbers around $0.2h$ Mpc$^{-1}$ \citep{Ouchi_2017,Herrera_2025, Yang_2009,Yang_2010,Ramakrishnan_2023, Barone_Nugent_2014, Harikane_2017, Hatfield_2018, Arita_2024, Dalmasso_2024, Paquereau_2025}.
In this work, we studied the non-Gaussianity of halo clustering at redshifts $z=8$ and $z=5$ and determined which terms in bias expansions are necessary for accurate modeling.
We found that bias terms that are powers of the matter density are more important than shape terms like the tidal tensor.
This is physically intuitive as these terms probe the densest regions where halos are expected to form.
This gave justification to expanding power spectrum and bispectrum models to higher orders in terms that are local in the linear matter density.
Our precise expansions are given in Eq. \ref{eq:power} for the power spectrum and Eq. \ref{eq:bispectrum} for the bispectrum.

We find that these bias expansions are able to fit the power spectrum and bispectrum at $z=8$ without giving any obvious systematic errors (Figures \ref{fig:Power_Plot} and \ref{fig:Bispec_Plot} respectively).
The power spectrum and bispectrum also independently give consistent constraints on bias coefficient values, shown on the left side of Figure \ref{fig:z8_Constraints}.

When fitting $z=8$ data, we confirm that bias terms which are local in matter density are more important than tidal terms.
All of our results are summarized in Table \ref{tab:1_sig_constraints}.
Even when using boxes of side length $2h^{-1}$Gpc at redshift $z=8$, we do not detect tidal bias at even a $1\sigma$ level.
In these same boxes, $b_3$ is detected at the approximately $1\sigma$ level. 
This shows that bias terms which are higher order in the matter field can be more important than lower order shape terms like the tidal tensor due to the much smaller values that tidal bias coefficients take.
This suggests that when expanding the halo density in terms of bias terms, the classic expansion treating $\delta$ as the perturbative parameter is not ideal, since we effectively get another perturbative parameter from the ratios of bias coefficients.
We additionally find that the nonlinearity of matter is detected at a higher significance than a tidal bias.
This can be understood with the same intuition as the preference of $b_3$ over tidal bias: nonlinearity of matter traces higher densities instead of shapes.

When looking at a slightly lower redshift of $z=5$ we find that this general behavior persists but becomes less pronounced.
Once again looking at Table \ref{tab:1_sig_constraints} we can see that at $z=5$, $b_3$ is detected at the $2\sigma$ level but a tidal bias is now detected at the $1\sigma$ level.
While $b_3$ is still detected with greater significance, the margin is more narrow than at $z=8$.
When comparing bias coefficients at different redshifts for catalogs with fixed variance, we find that the bias coefficients at $z=5$ are very nearly the bias coefficients at $z=8$ after applying linear evolution.
This effect can be seen in the right plot of Figure \ref{fig:Change_Constraints}.
This result suggests that the two halo samples were largely the same and the lower bias at $z=5$ can be attributed to the evolution of the matter density field.

While all of our work deals with the bias of halos, a natural generalization would be to apply this framework to galaxies produced in hydrodynamical simulations such as MillenniumTNG \citep{MillenniumTNG}.
The production of bias functionals done in this work is relatively cheap, and once produced they can be applied to any biased tracer at any redshift (after applying the growth factor).
To create more realistic observables, future works could additionally work in redshift space instead of real space.

While we do not work with galaxies or in redshift space, our general results still show the need for alternative bias expansions to properly analyze high-redshift tracers.
When tracers lie in the most overdense regions the traditional bias expansion breaks as there are significant differences in the values of bias coefficients.
In the high-redshift regime, tidal or shape-related contributions become subdominant and should be excluded for efficient descriptions. 
In contrast, higher-order density terms remain important and should be incorporated to accurately describe the bias.
Even with $2h^{-1}$Gpc simulation boxes, bias terms beyond $b_1$ and $b_2$ are never detected at more than a $2\sigma$ level.
Therefore, for future high-redshift galaxy surveys, bias expansions that consist of only $b_1$ and $b_2$ and using only the linear matter field would sufficiently capture clustering information.

\section{Acknowledgements}
\label{sec:Acknowledgements}

We thank Lehman Garrison for useful conversations.
KB was supported by the U.S.\ Department of Energy
Grant No. DE-SC0025671.  DJE was supported by 
U.S.\ Department of Energy grant DE-SC0007881 and as
a Simons Foundation Investigator.

\clearpage

\onecolumngrid

\bibliographystyle{apsrev4-2_16.bst}
\bibliography{main}

\end{document}